\documentclass[aps,floats,floatfix,amssymb,amsmath,prb,twocolumn,superscriptaddress]{revtex4-2}
\usepackage[english]{babel}
\usepackage{amsmath}
\usepackage{bm}
\usepackage{graphicx,bbm} 
\usepackage{times}
\usepackage{epsfig} 
\usepackage{soul}
\usepackage[utf8]{inputenc}
\usepackage[colorlinks,linkcolor=blue,citecolor=blue,urlcolor=blue]{hyperref}
\usepackage{color}
\usepackage{latexsym}
\usepackage{ulem}
\definecolor{nred} {RGB}{224,0,0}
\definecolor{nblue} {RGB}{28,130,185}
\definecolor{pgreen}{RGB}{78,138,21}
\definecolor{norange}{RGB}{230,120,20}

\begin{document}

\title{Easy-axis Heisenberg model on the triangular lattice: from supersolid to gapped solid}  
\author{M. Ulaga}
\affiliation{Jo\v zef Stefan Institute, SI-1000 Ljubljana, Slovenia}
\affiliation{\it Max Planck Institute for Physics of Complex Systems, Dresden, Germany}
\author{J.  Kokalj}
\affiliation{Faculty of Civil and Geodetic Engineering, University of Ljubljana, SI-1000 Ljubljana, Slovenia}
\affiliation{Jo\v zef Stefan Institute, SI-1000 Ljubljana, Slovenia}
\author{T. Tohyama}
\affiliation{Department of Applied Physics, Tokyo University of Science, Tokyo 125-8585, Japan}
\author{P. Prelov\v{s}ek}
\affiliation{Jo\v zef Stefan Institute, SI-1000 Ljubljana, Slovenia}

\begin{abstract}
We investigate the easy-axis Heisenberg model on the triangular
lattice by numerically studying excitations and the dynamical spin structure
factor $S^{\mu\mu}({\bf q},\omega)$. Results are analyzed within the supersolid scenario,
characterized by the translation-symmetry-breaking parameter $m_z$ and
the supersolid offdiagonal order parameter $m_\perp$. We find very robust $m_z >
0$ in the whole easy-axis anisotropy regime $\alpha = J_\perp/J_z >
0$, even enhanced by the magnetic field $h>0$, as well as
$m_\perp >0$ for intermediate $\alpha <1$ and $h>0$. Still,
at small $\alpha \lesssim 0.2$, relevant for recent experiments on the magnetic
material K$_2$Co(SeO$_3$)$_2$, we find at $h=0$ rather vanishing
$m_\perp \sim 0$, which appears compatible with the numerically established finite
magnon excitation gap $\Delta_1 \sim 0.25 \alpha J$.
       
\end{abstract}

\maketitle

\section {Introduction}

The antiferromagnetic (AFM)  Heisenberg spin-$1/2$ model on the
triangular lattice (TL) has been the origin of several fundamental
scenarios since its solution in the Ising limit revealed finite
entropy even at $T=0$ \cite{wannier50}. On the other hand, the
isotropic case has been the first candidate for the quantum spin
liquid (QSL) \cite{anderson73}, while later numerical studies
established the ground state (gs) as a symmetry-broken state, 
breaking translational symmetry with a $\sqrt{3} \times \sqrt{3}$ supercell and spins in $120^0$ alignment
\cite{bernu94,capriotti99,white07,chernyshev09}.  The span of
easy-axis anisotropies $0 <\alpha = J_\perp/J_z < 1$ opens another
interesting dimension \cite{miyashita85}. Whereas the gs
broken translational symmetry persists in the whole intermediate 
range $0< \alpha \leq 1$ (representing a spin solid with
longitudinal order parameter $m_z >0$), the most appealing is the 
scenario of a spin supersolid \cite{boninsegni12} which requires
simultaneously broken rotational in-plane symmetry and finite off-diagonal long-range order 
(LRO) signalled by $m_\perp >0$. Several 
numerical studies seem to confirm this possibility for $\alpha <1$
\cite{heidarian05,boninsegni05,wessel05,wang09,jiang09,yamamoto14,sellmann15},
leaving the question of its persistence for small $\alpha \ll 1$. 

The challenges revived with recent synthesis and experiments on several novel 
materials which represent the realization of the easy-axis Heisenberg  spin-$1/2$ model on TL.
The most interesting candidate is K$_2$Co(SeO$_3$)$_2$ (KCSO) \cite{zhong20}
which (due to the convenient value of $J_z$) allows for various experimental
investigations, in particular of thermodynamic quantities and spin excitation spectra 
via the inelastic neutron scattering (INS), in a wide range of temperatures $T$ and external fields $h$ 
\cite{zhu24,chen24} (see also \cite{mila24}).
Since the material is close to the Ising limit, i.e., with 
effective $\alpha \sim 0.07 $, the central question is whether it is in fact the realization of the spin  
supersolid. There are also other  challenging  novel materials, e.g.,
Na$_2$BaCo(PO$_4$)$_2$ \cite{li20,gao22,xiang24,gao24,sheng2025} with
$\alpha \sim 0.6$, 
closer to the isotropic case, and NdTa$_7$O$  _{19}$ \cite{arh22} with $\alpha \ll 1$,
which has so far experimentally revealed features closer to  QSL. 

We present results of  numerical finite-size studies of the easy-axis Heisenberg model on 
TL, which are consistent with  the supersolid scenario for intermediate $0.5 \le
\alpha <1$, but as well as at finite 
fields $h>0$ for the KSCO-relevant regime $\alpha \sim 0.1$. Still at $h \sim 0$
we find for small $\alpha \lesssim 0.2$,  
besides the robust quasi-elastic peak representing  diagonal LRO and finite $m_z >0$,
a rather vanishing $m_\perp \sim 0$ which is compatible with a finite magnon excitation gaps,
also established numerically.
This finding is consistent with our recent general study of thermodynamic properties of the model
\cite{ulaga24}  which were interpreted as a crossover/transition 
at $\alpha^* \sim 0.3$ to a regime $\alpha < \alpha^*$ characterized by finite excitation gaps.

In the Ising-like regime with $\alpha \ll 1$, it is very instructive to study static and dynamical
properties of the spin system, allowing to start the analysis from the extended magnetization 
plateau $m=1/3$ at finite $h \sim h_c$ (see also \cite{xu24}). By
decreasing $h<h_c$ our results
for the $T=0$ dynamical spin structure factor (DSSF) $S({\bf q},\omega)$,
calculated on finite systems via exact diagonalization (ED) up to $N = 36$ sites,
reveal a gapless magnon mode emerging from $m_\perp >0$, but also squeezed 
low-$\omega$ spectra originating from strongly correlated magnons. Still, on
approaching $h \sim 0$, the excitations appear to reveal a finite magnon gap
$\Delta_1 \propto \alpha J$ \cite{ulaga24}, which we confirm by the 
density-matrix renormalization group (DMRG) calculation on $N \leq 60$ sites. Moreover,
a similar conclusion that excitations might be anomalous follows also from a reduced effective
model, where translation symmetry is explicitly broken.

\section{Model}
 
We consider the anisotropic $S=1/2$ Heisenberg model with the nearest-neighbor (nn) exchange 
interaction $J_z =J$ and the easy-axis anisotropy $0 < \alpha \le 1$
on TL in the presence of a longitudinal  
magnetic field $h$,
\begin{equation}
H= J  \sum_{\langle ij \rangle} \lbrack S^z_i S^z_j + \frac{\alpha}{2}( S^+_i S^-_j  + S^-_i S^+_j)  \rbrack
- h\sum_{i}  S^z_i~~,~~\label{his}
\end{equation}
where the first sum runs over nn pairs on TL. Note that we further represent $h$ in units $J$.
Our previous study \cite{ulaga24} of finite-$T$ properties of the same model on TL, Eq.~(\ref{his}), 
employing the finite-temperature Lanczos method (FTLM) \cite{jaklic94,jaklic00}, 
already pointed out some results now directly relevant for KSCO. In particular, the
specific heat $c(T)$ exhibits a pronounced Schottky-like 
peak at $T^* \sim 0.3\alpha J$ for small $\alpha <\alpha^*$, which is well consistent with the experimentally 
observed $T^* \sim 1 K$
in KSCO \cite{zhong20} and estimated $J \sim 3\,$meV and $\alpha = J_\perp/J_z \sim 0.07$
\cite{chen24,zhu24}. Related is also large remanent spin entropy in KSCO at 
$T >T^*$ \cite{zhong20}. Results for the static spin structure factor 
$S^{zz}_{\bf q}(T \sim 0)$ \cite{ulaga24} also confirm the robust diagonal LRO at $T \sim 0$
consistent with a $\sqrt{3} \times \sqrt{3} $ spin solid.

In this work we focus on the $T=0$ (gs) DSSF $S^{\mu \mu}({\bf q},\omega) =
\langle \psi_0| S^\mu_{-\bf q} \delta(\omega - H + E_0) S^\mu_{\bf q}| \psi_0 \rangle$, 
with respect to the gs $| \psi_0 \rangle $ and its energy $E_0$ (in general 
for $h \ge  0$),  whereby $S^\mu_{\bf q}= N^{-1/2} \sum _i e^{i {\bf q} \cdot {\bf R}_i} S^\mu_i $ and 
$\mu =x,z$.  DSSF is calculated numerically  via ED, employing the Lanczos technique (see, e.g., Ref. \onlinecite{prelovsek13}), 
on TL  with $N=18 - 36$ sites  with periodic boundary conditions (PBC) for the related discrete 
${\bf q}$ in the Brillouin zone (BZ). 
It should be stressed that the application
of PBC (as well as finite-size scaling of results) appears crucial for such frustrated systems.

\section{Diagonal and off-diagonal long-range order}

\begin{figure}[t]
\centering
\includegraphics[width=\columnwidth]{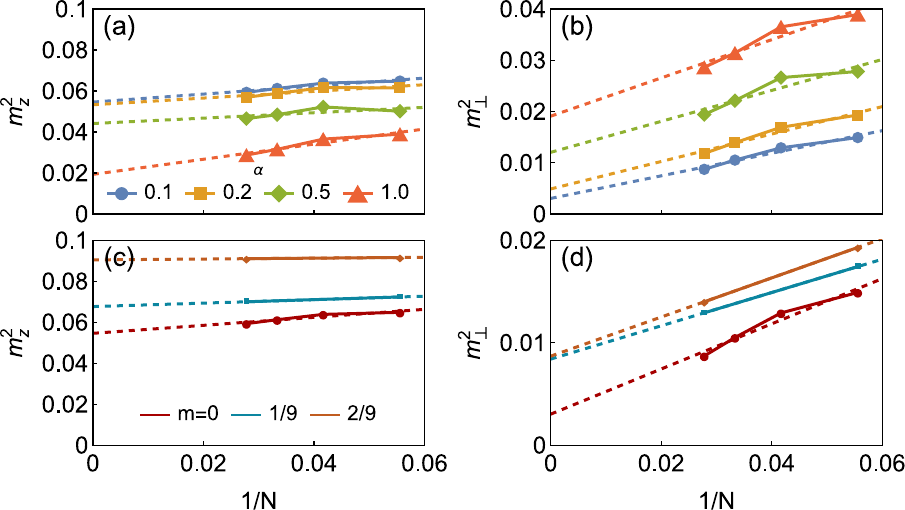}
\caption{(a) Diagonal LRO moments $m_z^2$ and (b) the off-diagonal $m_\perp^2$
vs. $1/N$ for different $\alpha = 0.1 - 1$ and $h=0$ as extracted from numerical results 
for $S^{\mu\mu}({\bf q}_K,\omega)$  on TL with $N=18 - 36$.  (c) $m_z^2$ and (d) $m_\perp^2$, again vs. $1/N$, 
for  $\alpha =0.1 $, but for $h \ge 0$ and corresponding magnetizations $m \ge 0$.}  \label{fig1}
\end{figure}

We first discuss the situation at $h=0$, where  the gs corresponds to $S^z_{\mathrm{tot}}=0$
and ${\bf q}_\Gamma =0$. The DSSF reveals a well-pronounced low-$\omega$
BZ corner mode at ${\bf q}_K = (4\pi/3,0)$ (in r.l.u.),  i.e., 
$S^{zz}({\bf q}_K,\omega) \sim A^{zz} \delta (\omega - \Delta_{0K})$ (excitations within 
the same $S^z_{\mathrm{tot}}$ sector) and  
$S^{xx}({\bf q}_K,\omega) \sim A^{xx} \delta (\omega - \Delta_{1K})$ (representing transitions with
$\Delta S^z_{\mathrm{tot}} = \pm 1)$, respectively.
The effective LRO parameters are then extracted  as $m_z^2 = A^{zz}/N$ and $m_\perp^2 = A^{xx}/N$, 
following their  asymptotic behavior for $N \to \infty$. Results, obtained for a wide range of $\alpha = 0.1 - 1$ 
and for systems with $N=18,24,30,36$ sites,
are presented in Fig.~\ref{fig1}(a,b). It should be acknowledged that, in principle, finite-size scaling of results 
(presented vs. $1/N$) comes with a caveat
since the considered systems (with PBC) have slightly different shapes (see also a  recent detailed analysis 
presented in Ref. \onlinecite{huang24}),
although they are all chosen to include the relevant ${\bf q}_K$ in the BZ. Some cluster-shape 
dependence  is  discussed in Appendix A.

Results in Fig.~\ref{fig1}(a,b) reveal qualitative differences
between nearly isotropic $\alpha \sim 1$ and the Ising-like $\alpha < \alpha^* \sim 0.3$ regimes
\cite{ulaga24}. In the isotropic case $\alpha =1$, the 
extrapolation for $N\to \infty$ yields consistent (and numerically nontrivial) $m_\perp = m_z \sim 0.14$.
By reducing $\alpha <1$  we establish increasing $m_z$ and decreasing $m_\perp$. 
While for $\alpha \le 0.2$ our results confirm the saturation of $m_z^2 \sim 0.06$ \cite{jiang09,xu24},
the asymptotic off-diagonal value is very small $m_\perp^2 \ll 0.01$, essentially too small for reliable
extrapolation. Moreover, the 
observed scaling $m_\perp^2 \propto 1/N$ implies that  $A^{xx}$ is rather $N$-independent, 
pointing to a finite magnon gap $\Delta_1 >0$, analyzed in more detail below.

\section{Excitation gaps}
  
A complementary message arises from the consideration of the lowest spin excitations.
For sizes $N \le 36$ we calculate them via ED directly or extract them from corresponding DSSF. 
Moreover, we employ here also the DMRG approach \cite{white93}, which allows to establish gs and 
first excited states in different $S^z_{\mathrm{tot}}$ sector, for much larger lattices $N \le 60$, again with PBC. 
The details on the method and corresponding results are presented in Appendix B.
In Fig.~\ref{fig2}(a) we show the evolution 
of the ``singlet'' gap $\Delta_0 = E^0_{{\bf q}_K} - E^0_0$ vs. $1/N$ from ED  ($E^0_0$ corresponding to 
${\bf q}_\Gamma = 0$), combined with the DMRG results for $\alpha \le 0.5$
(where values are well converged). The normalized $\Delta_0/(\alpha J)$ in Fig.~\ref{fig2}(a)
indicates a quantitative similarity for all $\alpha \le 1$, with (linearly in $1/N$) 
vanishing  $\Delta_0 \to 0$ for $N \to \infty$. This is consistent with diagonal LRO
at $N \to \infty$ with emergent $m_z > 0$ and a $\sqrt{3} \times \sqrt{3}$ supercell.  

\begin{figure}[t]
\centering
\includegraphics[width=1.0\columnwidth]{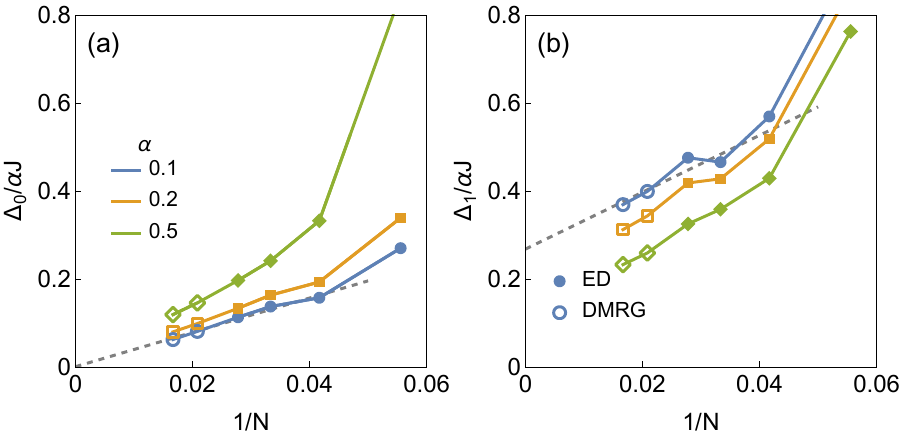}
\caption{ The normalized excitation gaps vs. $1/N$,  obtained from ED (for $N \le 36$) and DMRG (for $N > 36$): 
(a)  $\Delta S^z_{\mathrm{tot}} = 0$  ``singlet'' gap $\Delta_0/(\alpha J)$,
(b) $\Delta S^z_{\mathrm{tot}} =\pm 1$ magnon gap  $\Delta_1/(\alpha J)$, together with simple $N \to \infty$ extrapolations
for $\alpha = 0.1$.  }
\label{fig2}
\end{figure}

However, this is not the case for $\Delta_1$, representing the $\Delta S^z_{\mathrm{tot}} = \pm 1$ ``magnon'' gaps,
extracted from ED results as $\Delta_1 = E^1_{{\bf q}_\Gamma} - E^0_0$. We also note
that $\Delta_{1K} = E^1_{{\bf q}_K} - E^0_0 \sim \Delta_1 + \Delta_0 $,
as relevant for DSSF shown further on, should be the same in the limit $\Delta_0 \to 0$. 
In Fig.~\ref{fig2}(b) we  show $\Delta_1/(\alpha J)$ vs. $1/N$, as obtained via ED  and DMRG.
At least for $\alpha = 0.5$, $\Delta_1$ decreases with $N$, presumably consistent with  $m_\perp >0$. 
On the other hand, for $\alpha \le 0.2$, results support an  asymptotically finite
$\Delta_1 \sim 0.25 \alpha J$ (somewhat below the result in \cite{ulaga24}), consistent with the saturation of 
$A^{xx}$ and vanishing  $m_\perp$ in Fig.~\ref{fig1}(b). We note that the results for $\alpha = 0.5$ 
could admit a small but finite extrapolated value $\Delta_1 > 0$, which would be
inconsistent with $m_\perp >0$ in Fig.~\ref{fig1}(b), but this point evidently requires further
numerical efforts.

 \subsection{Finite fields $h>0$}
  
The question of the supersolid can be extended to finite fields 
$h >0$ and corresponding gs magnetizations $m = 2 S^z_{\mathrm{tot}}/ N > 0$, as directly relevant for 
experiments on KSCO \cite{zhu24,chen24}. In the regime of small $\alpha < \alpha^*$ there
is well pronounced $m=1/3$ plateau \cite{honecker04,ulaga24,xu24}, with gs $S^z_{\mathrm{tot}}=N/6$. 
For $\alpha \ll1$ the plateau appears at $h > h_c \sim 1.5 \alpha J$.  We perform the
calculation of  $S^{\mu \mu}({\bf q},\omega)$ for $m >0$  by choosing  proper $h>0$, and 
we repeat the analysis of $m_z$ and $m_\perp$ for $h < h_c$, i.e., $m< 1/3$.  
Since the same (commensurate) $m>0$ are allowed only is some lattices, we restrict  in Fig.~\ref{fig1}(c,d)  
results to $N=18,36$ systems (and $\alpha =0.1$).
We note that with increasing $m \rightarrow 1/3$ diagonal $m_z^2$ is even increasing, 
i.e., for $\alpha \ll 1$ towards the classical value  $m_z^2 \sim 1/12$. At the same time, 
the $0 < m < 1/3$ results indicate finite  $m_\perp > 0$ when 
extrapolated to $N\to \infty$, clearly in contrast to the $m=0$ case. It is also remarkable that the extracted 
$m_\perp$ (for $m>0$) are essentially $\alpha$ independent (not presented) for $\alpha \le 0.2$.

\begin{figure}[t]
\centering
\includegraphics[width=0.8 \columnwidth]{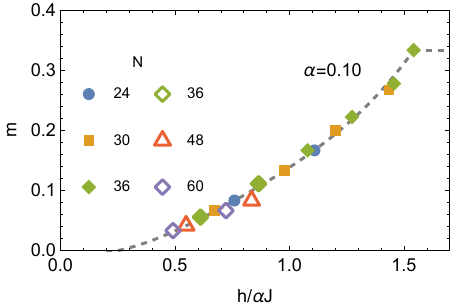}
\caption{Magnetization $m$ vs. renormalized magnetic field $h/\alpha$ at $T=0$
and $\alpha =0.1$, as calculated
from ED (full) and DMRG (empty symbols) gs energies in each $S^z_{\mathrm{tot}}$ sector. 
The line represents a simple interpolation targeted at $h \to 0$.}  
\label{fig3}
\end{figure}

The persistence of magnon gap $\Delta_1 >0$  has implications for the $T=0$ magnetization curve 
$m(h)$, in particular to the variation for $h \to 0$. 
With known gs energies $E^0_k$ within each spin sector $k = S^z_{\mathrm{tot}}/N = [- N/2, N/2]$
we can establish the $T=0$ magnetization curve $m(h)$ by 
using here the interpolation $h = (E^0_{k+1} - E^0_{k-1} )/2$, whereby
the related magnetization is $m =  k N/2$. We are interested in the regime 
below the plateau $m \le 1/3$. Besides ED results for $N \le 36$ available for all 
$k \le N/2$ we take into account also  DMRG results, presented in Appendix B and Fig.~\ref{fig8}, which
are crucial in the most delicate regime $ h \to 0$. In Fig.~\ref{fig3} we present the 
normalized magnetization curve $m(h/\alpha)$ for chosen $\alpha =0.1$,
which is (within our finite-size restrictions) nearly identical to the result
for $\alpha =0.2$. While such curves have been studied and presented for the 
isotropic case $\alpha \sim 1$ \cite{honecker04}, mostly discussed in relation 
to the existence and vicinity of the plateau regime, here we focus on the
weak fields $h \to 0$. It is evident that finite $\Delta_1 >0$ has qualitative consequences 
and should lead finally to vanishing $m(h<\Delta_1)=0$, and this 
tendency is observable in Fig.~\ref{fig3}. It should be, however, pointed that 
(when compared to experiment, e.g. \cite{zhu24}) the variation at $h \to 0$ is very sensitive 
to temperature $T$, requiring at least $T < \Delta_1$.  

\subsection{Effective model}

The observation that for $\alpha \ll 1$ the diagonal LRO $m^2_z \sim 1/12$ 
in the whole range $m < 1/3$, gives the justification to consider a
reduced spin model, where in Eq.~(\ref{his}) we explicitly break the translational
symmetry and fix spins on one sublattice to $S^z_i=-1/2$.  This gives an anisotropic $\alpha \ll1$ model, still Eq.~(\ref{his}), 
but now effectively on a 
honeycomb lattice (HL) and with fields $\tilde h = h + 3J/2$. Such a model remains nontrivial 
due to strong correlations  (at $\alpha \ll 1$) between remaining spins. But at least it allows
numerical consideration of larger lattices, in particular, a more detailed evolution
starting from the $m=-1/3  + 2 \tilde m/3$ (TL) plateau, $\tilde m$ representing
the effective magnetization in HL.

\begin{figure}[t]
\centering
\includegraphics[width=0.99\columnwidth]{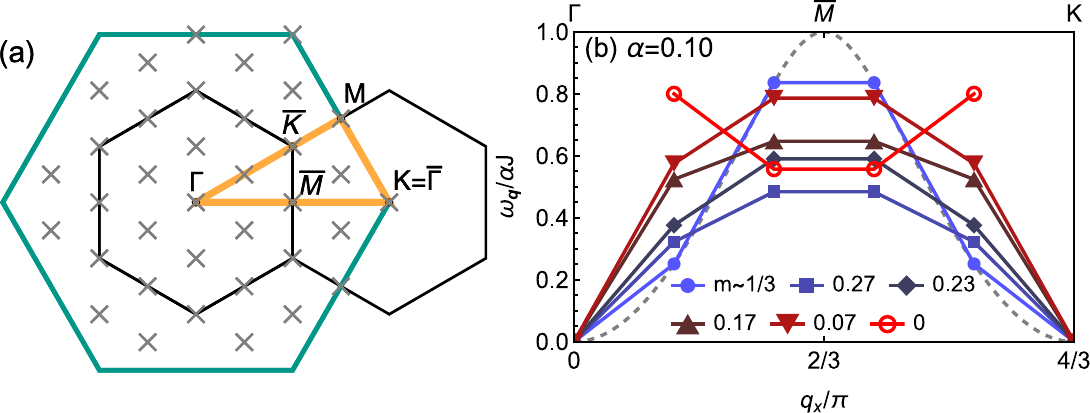}
\caption{(a) The Brillouin zone for TL (green) and the reduced
 Brillouin zone for the effective model on HL (black) with marked high-symmetry points,
 together with discrete ${\bf q}$ in the BZ for TL on $N=36$ sites,
 (b) Lowest spin excitations $\omega_{\bf q}/(\alpha J) $ within the same $S^z_{\mathrm{tot}}$ sector for 
different $m$ in the reduced model, as calculated numerically for $\alpha =0.1$ 
on TL with $N=60$ sites. The dispersion relation in Eq.~\ref{eff} is indicated with a dashed line.}  \label{fig4}
\end{figure}

The model at $\tilde m \lesssim 1$, just below the plateau $m \lesssim 1/3$, is
also solvable using magnon excitations with the dispersion (in r.l.u. of TL),
\begin{equation}
\omega^\pm_{\bf q}= \frac{\alpha J}{2} [3 \pm | f_{\bf q}|], \quad f_{\bf q} = e^{i q_x} 
+ 2  e^{- \frac{i}{2} q_x} \cos(\frac{\sqrt{3}}{2} q_y), \label{eff}
\end{equation}
These excitations have two branches, with a Dirac-like point $\omega^+_{{\bf q}_{\tilde K}}=
\omega^-_{{\bf q}_{\tilde K}}$ along the $\Gamma-M$ line 
(at the corner $\Bar{K}$  of the HL BZ, Fig.~\ref{fig4}(a)), gapless excitations 
$\omega^-_{\bf q} \propto q^2$, and $\omega^-_{\bf q} \propto \tilde q^2,~\tilde {\bf q}= 
{\bf q} - {\bf q}_K$. Such dispersion should well represent 
the spin-excitation spectra of the full model at $h \sim h_c$ as well
also experimental INS results in KSCO close to  $h \lesssim h_c$ \cite{zhu24}.

Still, it is challenging to determine the evolution of low-$\omega$ spin excitation 
in the effective model,  when increasing $\tilde m \to 1/2$, i.e, reducing $m \to 0$. 
We present in Fig.~\ref{fig4}(b) the numerical result for the dispersion of the lowest excitations
$\omega_{\bf q} = E_{\bf q} -  E_{0}$ within the same $S^z_{\mathrm{tot}}$ sector for different effective $m$, as obtained
now with ED on the largest system corresponding to TL on $N=60$ sites. While for  
$m \sim 1/3$ results agree with the analytical $\omega^-_{\bf q}$, Eq.~(\ref{eff}), 
the dispersion squeezes (relative to $m \sim 1/3$) as well as becomes supersolid-like 
for small  $q, \tilde q \to 0$ with decreasing $m$. Still, on approaching 
$m  \sim 0$ there is evident qualitative change and the results in Fig.~\ref{fig4}(b) for
$m = 0$ reveal anomalous excitation spectra  where again
$\omega_{{\bf q} \ne 0} > \alpha J/2$ \cite{ulaga24}. This is in contrast 
to the behavior of the dispersion in the effective model at large $\alpha \geq 0.5$
(see Appendix  C) where the anomaly visible at $m = 0$ is less pronounced. Still, the similarity
(or difference) of the effective model deserve more study.

\section{Dynamical spin response}

Let us turn to more complete results for the gs $S^{\mu \mu}({\bf q},\omega)$, 
focusing on the Ising-like regime. For $\alpha \ll 1$ $\mu\mu$ polarizations  
can be qualitatively different and partly complementary, since $zz$ response conserves 
$S^z_{\mathrm{tot}}$, while $xx$ component reflects $\Delta S^z_{\mathrm{tot}} = \pm 1$ transitions. 
INS experiments measure the spin polarization perpendicular to
in-plane ${\bf q}$, therefore we show the corresponding $S^\perp({\bf q},\omega) =
S^{zz}({\bf q},\omega) + S^{xx}({\bf q},\omega)$. We present results obtained via ED 
on the largest  TL with $N=36$ sites, which has rotational symmetry and contains the most relevant ${\bf q}$, 
in particular, BZ boundary ${\bf q}_K$ and ${\bf q}_M$.  Still, finite-size limitations
(also due to $T=0$ restriction) remain visible both in ${\bf q}$ as well in $\omega$ 
resolution.

\begin{figure}[t]
\centering
\includegraphics[width=\columnwidth]{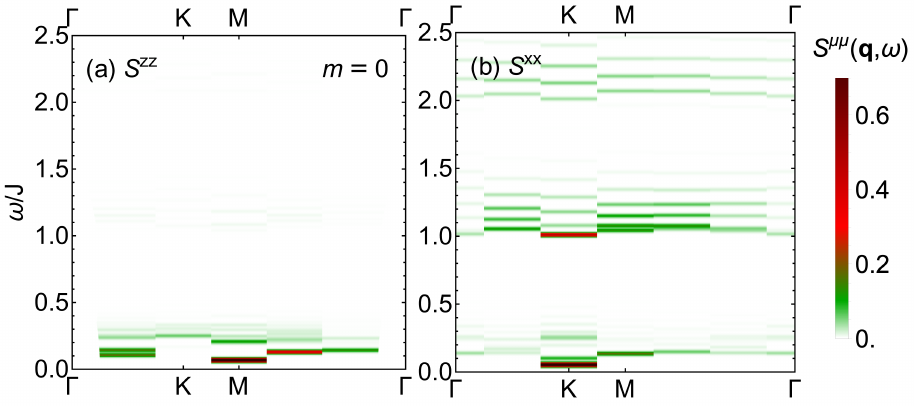}
\caption{ $T=0$ DSSF (a) $S^{zz}({\bf q},\omega)$  and (b) $S^{xx}({\bf q},\omega)$,
as obtained for $h=0$ and $\alpha =0.1$ on TL  with $N=36$ 
sites, presented in a broad $\omega/J \gg 1 $ range. Spectra are artificially broadened 
with broadening $\eta=0.01$.}  
\label{fig5}
\end{figure}

We present DSSF without the very strong quasielastic peak 
at $\omega = \Delta_0$, dominating $S^{zz}({\bf q}_K,\omega)$ and consequently 
the whole DSSF. Let us first comment on $S^\perp({\bf q},\omega)$ spectra 
for $h=0$ ($S^z_{\mathrm{tot}}=0$). Besides the most interesting dynamical regime 
$\omega <  3 \alpha J $ (discussed in detail further), there are also well-visible nearly
dispersionless excitations at $\omega \sim J$ and $\omega \sim 2J$,
which are also present in the INS results for KSCO \cite{chen24}.
Fig.~\ref{fig5} shows DSSF at $h = m=0$,  here separately for $S^{zz}({\bf q},\omega)$ and 
$S^{xx}({\bf q},\omega)$,  for chosen  $\alpha =0.1$ and ${\bf q}$ along the 
$\Gamma - K - M - \Gamma$ line in the BZ. Results, as obtained
for discrete ${\bf q}$, here via ED for TL with $N=36$ sites,  are extended in ${\bf q}$ to
improve visibility. While the lowest energy excitations are represented in both components,
the higher energy branches of nearly dispersion-less magnetic excitations 
appearing at $\omega \sim n J$ are well pronounced only in $S^{xx}({\bf q},\omega)$.  
They emerge from  spin flips with $\Delta S^z_{\mathrm{tot}}=\pm 1$
in local Ising-like spin nearest-neighbor environments $S^z_{\mathrm{loc}}= n$.

\begin{figure}[t]
\centering
\includegraphics[width=\columnwidth]{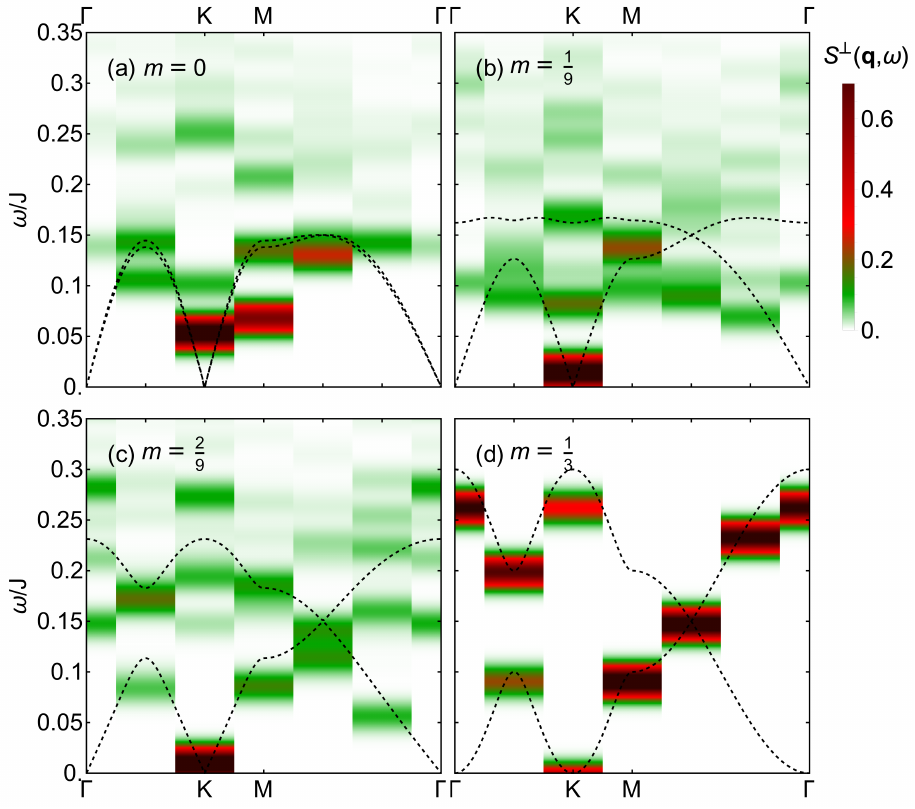}
\caption{ Low-$\omega$ regime of the gs DSSF 
$S^\perp({\bf q},\omega)$  for $\alpha =0.1$ and various
magnetizations $0 \le m \lesssim 1/3$ 
obtained on TL  with $N=36$  sites. Dashed black lines denote
 the LSW approximation dispersion for $m<1/3$ and the analytical
 result for the effective model at $m=1/3, h \sim h_c $. }
\label{fig6} 
\end{figure}

More challenging is the $\omega < \alpha J$ regime and its evolution 
with the field. The summary of INS-relevant 
$S^\perp({\bf q},\omega)$, as it develops in the sub-plateau regime $ 0 \le m < 1/3$,
is presented in Fig.~\ref{fig6} for fixed $\alpha =0.1$ and for 
${\bf q}$ along the  $\Gamma - K - M - \Gamma$ line in BZ (orange line
Fig.~\ref{fig4}(a)).   Since both ${\bf q}$ and $\omega$ spectra 
(at $T=0$) are discrete, results are represented as broadened for convenience. 
We start the interpretation with the simplest case, i.e., at the onset  
of the $m=1/3$ plateau at $h \lesssim h_c$ in  Fig.~\ref{fig6}(d). The magnon dispersion
here closely follows the analytical expression, Eq.~\ref{eff}, with both lower and upper 
branches $\omega^\pm_{\bf q}$ being sharp and well visible, with the
main contribution from  $S^{xx}({\bf q},\omega)$.

The evolution with decreasing but finite $0 < m < 1/3$ in   Fig.~\ref{fig6}(c,b) reveals
several generic features: (a) Spectra are less coherent, although with rather 
well-pronounced lower edges. (b) Consistent with the concept of supersolid and $m_\perp >0$,
the spectra are (nearly) gapless at ${\bf q}_K$, whereby 
the main contribution emerges from the $S^{xx}$ component. (c) The whole spectra 
still partly reflect two branches, but are effectively squeezed in
$\omega$ relative to the $m \sim 1/3$ case.
(d) A pronounced dynamical response at ${\bf q}_M$, originating from 
$S^{zz}({\bf q},\omega)$, moves down in $\omega$ with decreasing $m$ and becomes subdominant 
compared to the soft ${\bf q}_K$ peak in $S^{xx}({\bf q},\omega)$. While certain
features discussed above remain even for the most interesting $m=0$ in
Fig.~\ref{fig6}(a), there are some essential differences: (a) The lowest $q \ne 0$ and also 
the most pronounced  excitation emerging from  $S^{xx}({\bf q}_K,\omega)$
is now gapped, consistent with $\Delta_{1K} \propto \alpha J$ in Fig.~\ref{fig2}b, 
(b) there is also well-pronounced low-$\omega$  excitation at
${\bf  q}_M$ emerging from $S^{zz}({\bf q},\omega)$, consistent with INS
experiment on KSCO \cite{zhu24} and reminiscent of the roton-like 
minimum in the isotropic TL \cite{zheng06,ferrari19}.

\subsection{Linear spin-wave theory}
 
It is instructive to consider the linear spin wave (LSW)
approximation for the full model, as also partly shown
in Refs.~\onlinecite{zhu24,xu24}.  The LSW approximation starts from
classical ground state, which in the discussed regime has three
sublattices, with spin on one sublattice pointing down, while the other
two pointing up, but at certain angle $\vartheta$ (and symmetrically) from the $z$
direction (the planar ``Y'' state). The energy of such a state
can be expressed as 
\begin{equation}
\frac{E(\vartheta)}{N}=-S^2J(2\cos\vartheta -\cos^2\vartheta
+\alpha\sin^2\vartheta) -\frac{h S}{3}(2\cos\vartheta -1),
\end{equation}\noindent
with the ground state $\vartheta$ corresponding to the minimum of $E(\vartheta)$ . 
With decreasing $\alpha$ and increasing $h$ the angle $\vartheta$  is becoming smaller, which decreases
$m_\perp = S \sin\vartheta$, and can be expressed analytically as
\begin{equation}
m_\perp/S=\sqrt{1-[(1+h/(3SJ))/(1+\alpha)]^2)},
\end{equation}\noindent 
with $S=1/2$, i.e., $m_\perp \sim \sqrt{\alpha/2 - h/(3J)}$  for $\alpha \ll1$ relevant here.  
As $\alpha\to 0$, the classical result gives  decreasing $m_\perp\sim
\sqrt{\alpha/2}$.

Once the classical ground state is determined, the LSW dispersions are
calculated by following Ref. \onlinecite{toth15} and shown in Fig~\ref{fig6}.
We note that at $m\sim 1/3$ the full model LSW approximations
give slightly lower energies at the upper edge of the second branch, than
the effective HL model and the analytical result shown in Fig.~\ref{fig6}(d).
 The LSW results shown in the  in Figs.~\ref{fig6}(b,c) are
calculated at the magnetic fields $h$ that give the corresponding
magnetizations $m$ in the numerical ED calculation.  We also find that
for small $\alpha$ and $h=0$ the lower branch of the LSW dispersion possesses a
width and magnon velocity $v$ proportional to $\alpha$, while for $h>0$
it is well approximated by
$v(\alpha,h)\sim \alpha\,m_\perp(\alpha,h)/m_\perp(\alpha,h=0)$.

LSW at $h=m=0$ suggests
$m_\perp\sim \sqrt{\alpha/2}$ for $\alpha \to 0$,
decreasing with $h >0$ and vanishing at $h>h_c$. For $m>0$ this
qualitatively (but not quantitatively) agrees with the numerical result in
Fig.~\ref{fig1}(b), but clearly  disagrees with vanishing $m_\perp \sim 0$
at $h \to 0$. Finally, for  $m\sim 1/3$ and $m=2/9$ shown in Figs. \ref{fig6}(c,d) the maximal spectral
intensity resembles qualitatively the LSW dispersion, while for
$m=1/9$ and $m=0$, the agreement is much worse, with a much broader
numerical spectra and additional pronounced modes at ${\bf q}_K$ and
${\bf q}_M$.

\section{Discussion}

Our results confirm that at finite $h>0$, besides even increased longitudinal $m_z >0$,
there is firm evidence for off-diagonal LRO $m_\perp > 0$, consistent with the 
theoretical \cite{jiang09,xu24} and experimental \cite{zhu24,chen24} interpretation in 
terms of a spin supersolid. This is, however, not 
the case for the most challenging  $h \sim 0$ case, where our extrapolated 
ED results rather indicate (within numerical resolution) nearly vanishing $m_\perp \sim 0$,
which is at least in strong disagreement with the LSW approximation (having the largest 
$m_\perp$ at $h=0$). 
This finding is in fact not inconsistent with tiny (and extrapolation-limited) $m_\perp \gtrsim 0$
in previous studies \cite{jiang09} as well as very recent similar conclusions of 
Refs.~\onlinecite{xu24,gallegos2024}. 
Moreover, the observed $m^2_\perp \propto 1/N$ 
is compatible with the nonvanishing magnon excitation gap  $\Delta_1 \propto \alpha J$,
well visible in ED results for $S^{xx}({\bf q},\omega)$ and confirmed directly by DMRG results on 
systems with up to $N \le 60$ sites. The gap is also reflected 
in the absence of low excitations in specific heat $c(T<T^*)$  \cite{ulaga24}.
In this respect, the large anisotropy regime $\alpha \le 0.3$ \cite{ulaga24}
could be different from less anisotropic $\alpha \le 1$, where the spin supersolid
appears to be realized even at $h=0$ \cite{li20,gao22,gao24}. 
It should be acknowledged that the behavior at intermediate
$0.2 < \alpha \lesssim  0.7$ was not a focus of the present study and
still represents  a (in particular numerical) challenge. This includes the 
presumable transition/crossover at $\alpha \sim \alpha^*$  from gapped solid to
supersolid even at $h \sim 0$ (with previously estimated $\alpha^* \sim 0.3$ \cite{ulaga24}).

Making contact to experiments on KSCO, and taking into account 
assumed values for $J \sim 3\,$meV and $\alpha \sim 0.07$, our best DMRG estimate 
$\Delta_1 \sim 0.25 \alpha J$ would yield $\Delta_1 \sim 0.06\,$meV, which 
might be even compatible with recent INS spectra $S^\perp({\bf q},\omega)$ 
at ${\bf q} \sim {\bf q}_K$  \cite{zhu24,chen24}.  Otherwise, our calculated DSSF 
overall correspond well to INS results \cite{zhu24},
whereby lower branches can be partly captured by the LSW approximation.

We observe the gap also within a related effective model on HL
allowing ED to reach numerically larger systems.  It is evident that in this model magnetic excitations
for $\alpha \ll 1$ are strongly repulsive 
whereby the $h=m=0$ case corresponds to a commensurate filling in HL. 
So similarities can be found to the gapped magnon excitations
in dimerized $J_1$-$J_2$ chains \cite{majumdar69,uhrig96,pouget17} or the planar 
Shastry-Sutherland model \cite{shastry81}. Nevertheless, firm establishment and the understanding of 
the magnon gap as well as its presumable vanishing with increasing $\alpha$ remains a future challenge.
 
\section*{Acknowledgments}
 
We thank A. Zheludev and A. Zorko for stimulating discussion of recent experimental results. 
M.U. further acknowledges discussions with Alexander Wietek, Rafael Alvaro Flores Calderon and Roderich Moessner.
This work is supported by the program P1-0044 and project J1-50008 of the Slovenian 
Research Agency. M.U. acknowledges computing time at HPC 
Vega at the Institute of Information Science in Maribor under project S24O01-28.

\appendix

\section{Cluster shape analysis}

\begin{figure}[t]
  \centering
 \vspace{3mm}
\includegraphics[width=0.7\columnwidth]{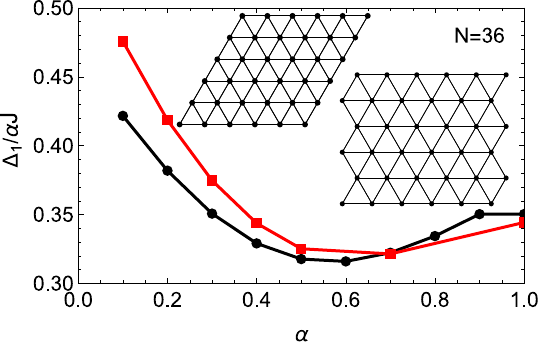}
\caption{The magnetic gap as a function of anisotropy on inequivalent
  $N=36$ clusters: the rhomb (red) and square (black).}   
\label{fig7}
\end{figure}

Some dependence of the cluster shape chosen for the $N=36$ system is summarized in Fig. \ref{fig7}. 
The rhombic cluster selected for the analysis in the rest of the manuscript is fully compatible with the 
infinite lattice symmetry, including 6-fold rotational symmetry, and possesses 7 inequivalent $\bf q$ points.
Alternatively, one can choose a ``square-ish'' cluster with lesser symmetry and more, 15, 
inequivalent $\bf q$ points. We note that the
magnetic gaps $\Delta_1$ differ by $\sim 10\%$ between the two
clusters with the gap being slightly smaller on the square cluster at
small  $\alpha$, but generally follow the same pattern. The shape
influence on nonmagnetic gaps $\Delta_0$ (not shown) is minimal. At the same time,
the calculated $m_\perp$ appear even less dependent on the shape,
i.e., on the $N=36$ cluster obtained values differ less than $1\%$ between rhombic and square shape.

\section{DMRG method and results}

In the DMRG calculations, we use $N=6\times6,$ $6\times 8$, and $6\times 10$ clusters of rhombic shape 
with PBC and perform a snake-type sweeping procedure. 
The bond dimension is taken to be $\chi=8000$, leading to a truncation error less than $2\times 10^{-5}$ for 
$\alpha=0.1$ and $\alpha=0.2$. This $\chi$ gives good convergence of the energy: for example, a relative 
energy difference between $\chi=5000$ and 8000 is less than 0.3\% for $\alpha=0.2$ at $N=60$. 
A decrease in accuracy for large 
$\alpha$ is due to the increase of transverse exchange terms proportional to $\alpha$, which 
inevitably induces effective long-range hopping in the sweeping process of DMRG under PBC. 

\begin{figure}[t]
\centering
\includegraphics[width=0.7\columnwidth]{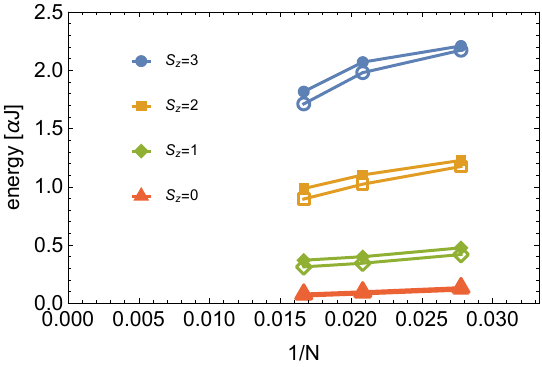}
\caption{Lowest excitation energies $E^0_k/ (\alpha J)$  for different $k=S^z_{\mathrm{tot}}/L$ sectors, 
relative to the gs $E^0_0$, as calculated via DMRG for different systems with $N =36 - 60$
sites, for two anisotropies: $\alpha =0.1$ (full symbols),  and $\alpha =0.2$ (empty).  }  \label{fig8}
\end{figure}

\begin{figure}[t]
\centering
\includegraphics[width=1.0\columnwidth]{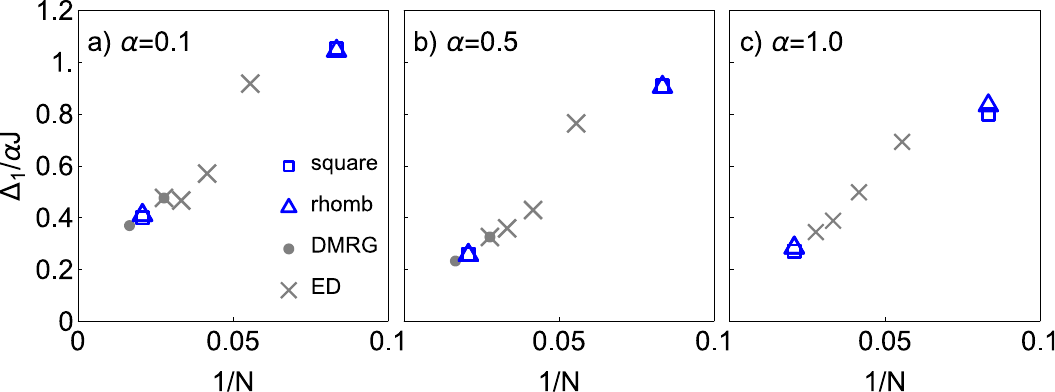}
\caption{Results for the normalized $\Delta_1$ gap, obtained via DMRG on lattices of 
$N= 12, 48$ sites with constant $L_x/L_y$ (in blue)  and their shape (rhombic vs. square) 
dependence for:  (a) $\alpha = 0.1$, (b) $\alpha = 0.2$, and (c) $\alpha = 1$, respectively. 
Shown are also other ED and DMRG results, obtained on lattices of different shapes.}  \label{fig9}
\end{figure}

While above considered DMRG calculations are performed on rhombic shapes $N =L_x \times L_y$ with various $L_x/L_y$,
one can test the same sizes also on nearly square-like shapes (see, e.g., Fig.~\ref{fig7}). In  
Fig.~\ref{fig9} we present DMRG results for $\Delta_1$, obtained  for different $\alpha =0.1, 0.5, 1.0$ 
on $N = 12, 48$ lattices with  rhombic and square-like shapes, together again with other ED and DMRG
on lattices with different shapes. The apparent near-independence on the shape leads to 
conclude that DMRG with PBC is less sensitive to lattice shapes and side ratios, but sizes evidently matter.

\section{Effective model: further results}

\begin{figure}[t]
\centering
\includegraphics[width=0.99\columnwidth]{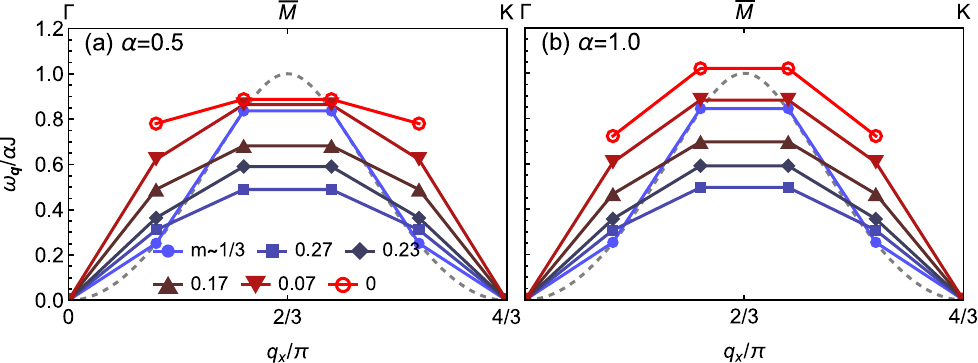}
\caption{ Lowest spin excitations $\omega_{\bf q}/(\alpha J) $ within the same $S^z_{\mathrm{tot}}$ sector
for  different $m$ in the reduced model, as calculated numerically on TL with $N=60$ sites 
for : (a) $\alpha =0.5$, and (b) $\alpha =1.0$. }  \label{fig10}
\end{figure}

The effective model, representing the anisotropic Heisenberg model on a honeycomb lattice, has been 
obtained from the full model on TL by freezing spins on one sublattice. While such reduction is well
justified in the case of strong anisotropy $\alpha \ll 1$, one can consider its behavior also more
generally with increasing $\alpha \lesssim 1$. We present (in analogy with Fig.~\ref{fig4}) in Fig.~\ref{fig10} 
the lowest  spin excitations $\omega_{\bf q}/(\alpha J) $ within the same $S^z_{\mathrm{tot}}$ sector for different $m$, 
but now calculated at $\alpha =0.5$ and $\alpha =1$. While for $ m > 0$ the results are even quantitatively 
similar for all $\alpha \leq 1$, this is evidently not the case for dispersion at $m=0$. In contrast 
to $\alpha = 0.5,\;1.0$ the behavior for $\alpha =0.1$ in Fig. ~\ref{fig4}(b) is anomalous
at  small $ q \to 0$. 

%

\end{document}